# On-Surface Pseudo-High Dilution Synthesis of Macrocycles: Principle and Mechanism

*Qitang Fan[1,2], Tao Wang[1], Jingya Dai[1], Julian Kuttner[2], Gerhard Hilt[2], J. Michael Gottfried[2]\*, Junfa Zhu[1]\**

[1]National Synchrotron Radiation Laboratory and Collaborative Innovation Center of Suzhou Nano Science and Technology, University of Science and Technology of China, Hefei 230029, P.R. China, jfzhu@ustc.edu.cn

[2]Fachbereich Chemie, Philipps-Universität Marburg, Hans-Meerwein-Str., 35032 Marburg, Germany, michael.gottfried@chemie.uni-marburg.de

**ABSTRACT**

Macrocycles have attracted much attention due to their specific "endless" topology, which results in extraordinary properties compared to related linear (open-chain) molecules. However, challenges still remain in their controlled synthesis with well-defined constitution and geometry. Here, we report the first successful application of the (pseudo-)high dilution method to the conditions of on-surface synthesis in ultrahigh vacuum (UHV). This approach leads to high yields (up to 84%) of cyclic hyperbenzene ([18]-honeycombene) *via* an Ullmann-type reaction from 4,4''-dibromo-*meta*-terphenyl (DMTP) as precursor on a Ag(111) surface. The mechanism of macrocycle formation was explored in detail using scanning tunneling microscopy (STM) and X-ray photoemission spectroscopy (XPS). We propose that hyperbenzene (MTP)$_6$ forms majorly by stepwise desilverization of an organometallic (MTP-Ag)$_6$ macrocycle, which preforms *via* cyclisation of (MTP-Ag)$_6$ chains under pseudo-high dilution condition. The high probability of cyclisation on the stage of the organometallic phase results from the reversibility of the C-Ag bond. The case is different from that in solution, in which cyclisation typically occurs on the stage of covalently bonded open-chain precursor. This difference in the cyclisation mechanism on a surface compared to that in solution stems mainly from the 2D confinement exerted by the surface template, which to a large extent prevents the flipping of chain segments necessary for cyclisation.

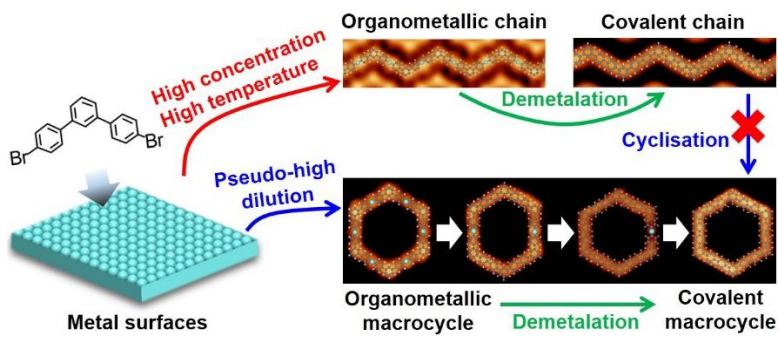

Tailoring the topology of polymers has intrigued chemists for many years, because a polymer's macroscopic properties depend inherently on its nanoscopic topology. Due to their specific "endless" topology, cyclic polymers bear significantly different characteristics compared to related linear (*i.e.*, open-chain) molecules, including higher density, lower intrinsic viscosity, and higher thermostability.[1-9] To date, the two most frequently used synthetic routes to cyclic polymers are high-dilution and molecular template methods performed in solution. The high-dilution principle, according to which low concentrations of the starting precursor favor cyclization over chain formation, was first developed by Paul Ruggli and Karl Ziegler for the cyclization of small organic molecules.[10,11] The template method was developed to guarantee that the target molecule has the desired shape[12] and thus was employed to synthesize various complicated macrocycles. An example for the latter is the recently reported Vernier templating synthesis of π-conjugated porphyrin nanorings with different diameters.[13-16] The main drawback of these solution-based methods is that the target cyclic polymer or macrocycles cannot be arbitrarily designed due to the solubility requirements of the reactants.[17] This drawback can be compensated by the on-surface synthetic approach under solvent-free conditions, which enable the polymerization and cyclisation of even insoluble reactive organic molecules directly on a solid surface.[17-21] As an additional advantage of the on-surface approach, the properties of the target macrocycles can be explored *in situ* with powerful surface science techniques, such as scanning tunneling microscopy/spectroscopy (STM/STS) and photoemission spectroscopy (PES). Recently, several macrocycles were successfully produced *via* the on-surface Ullmann reaction,[22] including honeycombenes,[17,18] sexiphenylene,[19] oligothiophene nanorings,[20] and the $(DAD)_6$ cycle with bis(5-bromo-2-thienyl)-benzobis(1,2,5-thiadiazole) as the precursor monomer.[21] Nevertheless, the yields of these macrocycles are still limited. This is mainly because chain-growth dominates over ring-closure under the typically employed reaction conditions on surfaces, *e.g.*, precursor deposition rate in the range of 1 to 100 monolayers per hour (ML/h).

To enhance the probability of ring-closure in the ring/chain competition, we have transferred the high-dilution principle, which is well proven for synthesis in solution, to the solvent-free conditions of on-surface synthesis. This principle uses the fact that cyclization of a chain is a first-order reaction, while the chain growth by attachment of another precursor is a second-order reaction. With decreasing concentration, the first-order cyclization competes more and more successfully against the second-order chain growth.[8] We show that rigorous application of the high-dilution principle to the synthesis of hyperbenzene ([18]-honeycombene) from 4,4''-dibromo-*meta*-terphenyl (DMTP) on Ag(111) results in considerably higher yield (84%) compared to that obtained under the usual conditions, *e.g.*, in previous works.[17, 18] In addition, the mechanism of the high-yield formation of hyperbenzene under pseudo-high dilution condition is revealed. It is found that the cyclisation happens on the stage of an intermediate organometallic phase, which reversibly forms (MTP-Ag)$_6$ macrocycles. These organometallic macrocycles then undergo stepwise demetalation forming hyperbenzene. This mechanism is different from the typical cyclisation mechanism in solution, which occurs on the stage of irreversibly (typically covalently) bonded open-chain precursors. Therefore, this study provides important general guidance for the high-yield on-surface synthesis of macrocycles.

**RESULTS AND DISCUSSION**

**On-surface pseudo-high dilution synthesis of hyperbenzene.** For an initial comparison between the pseudo-high dilution condition and typical high-concentration conditions, deposition of 0.6 monolayer (ML) 4,4''-dibromo-*meta*-terphenyl (DMTP) onto Ag(111) surface at 353 K was first performed with high flux (5 ML/h) to obtain a high initial concentration of the DMTP monomers. This leads to the formation of ordered islands of zigzag chains as shown in Figure 1a. The detailed structure of the zigzag chains is revealed by the magnified STM image in the inset, in which alternating corners and bright dot-like features are observed. According to our previous studies,[18, 23, 24] the corner motifs and bright dots are assigned to the *m*-terphenyl (MTP) units and Ag atoms, respectively, as illustrated by the superimposed molecular model in the inset of Figure 1a. In combination with the C-Br bond scission at 353 K, as evidenced by Br 3d XP spectra (Figure S1), the corner-to-corner distance of 16.0 Å strongly indicates, by bond length considerations, that C-Ag-C bonds are formed between the MTP units. This organometallic phase can be viewed as a reservoir of MTP precursors with high initial concentration. The organometallic chains can undergo transformation into hydrocarbon chains or macrocycles *via* elimination of the bridging Ag atoms and subsequent C-C coupling.

Subsequent annealing of the sample in Figure 1a to 463 K leads to the formation of both zigzag chains and hexagonal cycles as marked by the blue and red rectangles in Figure 1b, respectively. Their detailed structures are revealed in the zoom-in STM images in the inset of Figure 1b and 1c. For the zigzag chains, the corner-to-corner distance is reduced to 13.3 Å, indicating that C-C bonds are now formed between the MTP units (see the overlaid molecular model). A similar shrinking occurs for the hexagonal cycles in Figure 1c, in which the corner-to-corner distance is 13.1 Å. This suggests that hyperbenzene[18] (*i.e.*, [18]-honeycombene[17]) molecules are formed, as illustrated by the superimposed molecular model. Noteworthy, some of the hyperbenzene molecules have inclusions with bent and star-shaped features, which are assigned to single MTP units or diffusing Ag adatoms (see Figure S2 in the Supporting Information for structural details of the inclusions).

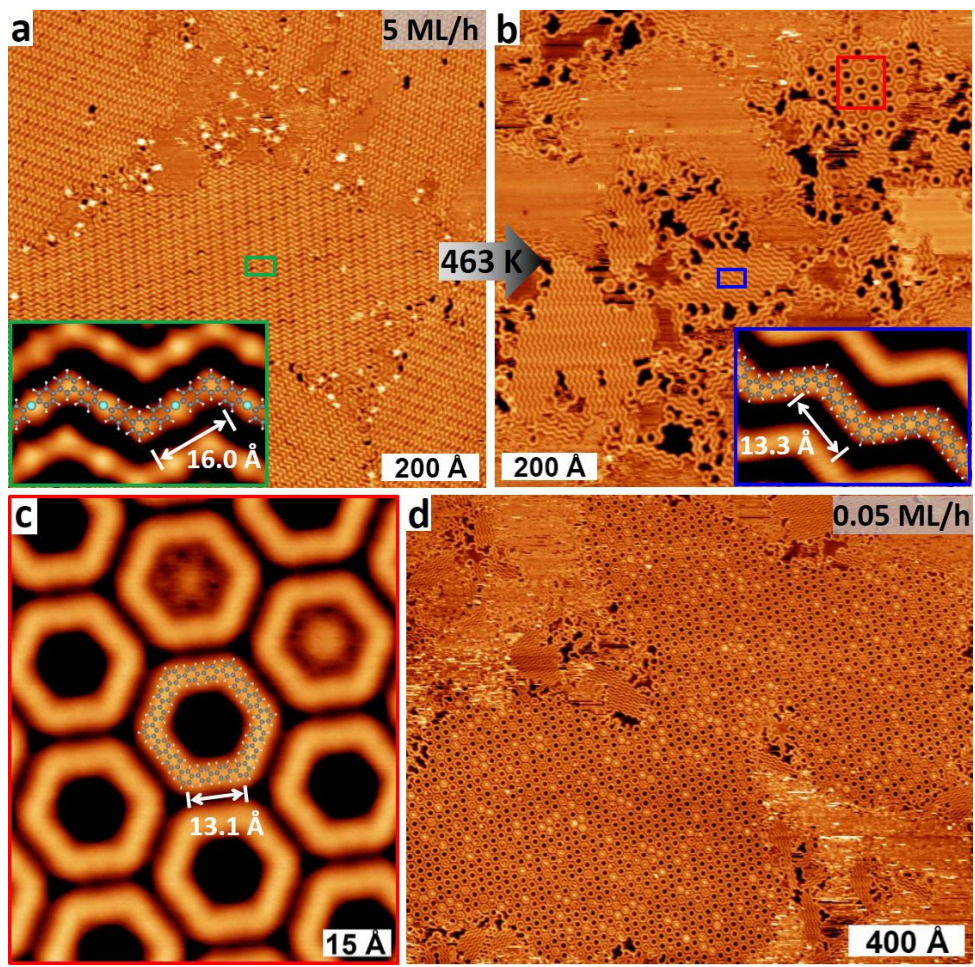

Figure 1. (a) STM image taken after deposition (flux $f$ = 5 ML/h) of 0.6 ML DMTP onto Ag(111) held at 353 K. The inset shows a magnified view of the green framed region. (b) STM image taken after annealing the sample in panel (a) to 463 K. The inset shows the blue framed region. (c) Zoom-in STM image of the red framed region in panel (b). Molecular models are overlaid. Grey spheres represent carbon atoms; white, hydrogen; cyan, silver. (d) Large-scale STM image taken after deposition of 0.5 ML DMTP onto Ag(111) held at 463 K with a deposition rate of 0.05 ML/h. Tunneling parameters: (a) $U$ = 1.4 V, $I$ = 0.08 nA; (b), (c) $U$ = 1.5 V, $I$ = 0.13 nA; (d) $U$ = 0.29 V, $I$ = 0.19 nA.

The formation of zigzag oligophenylene chains with high yield (94%) and hyperbenzene with low yield (6%) by post-annealing of the organometallic phase is typical for high-concentration conditions, and indeed the initial coverage of the MTP precursors (0.6 ML) is relatively high compared to the range that favors cyclization reaction: The two-dimensional (2D) coverage of 0.6 ML MTP corresponds to a three-dimensional (3D) concentration of approximately 0.6 mol/L. This value is much higher than the typical precursor concentrations of < $10^{-4}$ mol/L previously used for high-dilution cyclization reactions performed in solution.[25, 26] To obtain pseudo-high dilution conditions in the surface reaction, a very low rate of $f$ = 0.05 ML/h was now used during the deposition of DMTP onto the Ag(111) surface held at 463 K. In this way, the two main requirements for the pseudo-high dilution condition were satisfied:[27, 28] The first requirement (I) is that the reaction rate should be higher than the rate of addition of the precursor, such that the precursor concentration will not build up during the reaction. In fact, it can be estimated that the stationary concentration of DMTP is lower than $5 \times 10^{-4}$ ML for a flux of 0.05 ML/h (see part (d) in the SI for details). The second requirement (II) is that the reaction products should be stable under the reaction conditions. The requirement (I) was demonstrated by the fact that no excess of intact DMTP molecules exists, which was confirmed by STM and XPS measurements. The requirement (II) has been verified by previous studies on both Ag(111) and Au(111), in which the covalent products are very stable under the reaction conditions.[29-33] Remarkably, this procedure leads to a high yield (84%) of hyperbenzene macrocycles, as shown by the large-scale STM image (Figure 1d and Figure S3a$_2$) of the as-prepared sample. The result indicates that the pseudo-high dilution method on a surface indeed works similarly successful to that in solution.

In addition, the relationship between the yield of hyperbenzene and the deposition rate, $f$, of DMTP has been investigated in detail. Figure 2a-e shows overview STM images taken after deposition of 0.5 ML DMTP onto the Ag(111) surface at 463 K with different deposition rates of 0.08, 0.5, 5, 50, and 150 ML/h. The image series shows that the yield of hyperbenzene (green areas) decreases dramatically with increasing deposition rate $f$. Figure 2e plots the variation of the hyperbenzene yield with $f$, as derived from statistical analysis of the STM image series in Figure S3. The data points can be well fitted with an exponential curve (red) and illustrate that a low concentration of the DMTP precursors strongly favors the formation of hyperbenzene rings.

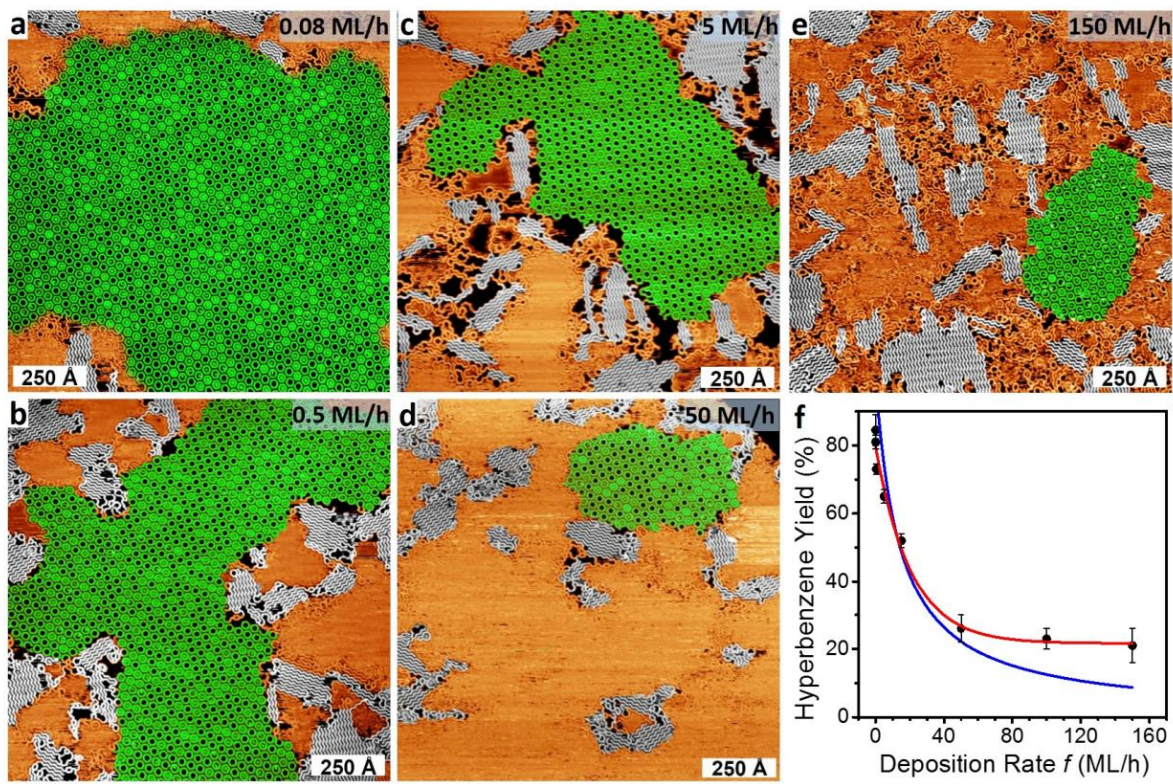

Figure 2. Overview STM image taken after deposition of 0.5 ML DMTP onto Ag(111) held at 463 K with different evaporation rates: (a) 0.08 ML/h, (b) 0.5 ML/h, (c) 5 ML/h, (d) 50 ML/h, and (e) 150 ML/h. Tunneling parameters: (a) $U = 1.2$ V, $I = 0.13$ nA; (b) $U = 1.3$ V, $I = 0.21$ nA; (c) $U = 1.6$ V, $I = 0.12$ nA; (d) $U = 1.5$ V, $I = 0.16$ nA; (e) $U = 1.6$ V, I = 0.13 nA. The islands of hyperbenzene and zigzag oligophenylene chains are indicated by green and grey shading, respectively. (f) Hyperbenzene yield versus DMTP deposition rate $f$. See the text for the meaning of the blue and red curves.

**Cyclisation on the stage of organometallic intermediates.** To obtain deeper insight into the mechanism of the hyperbenzene formation, the reaction temperature was lowered by 20 K to 443 K during the deposition of DMTP. In this way, we were able to observe intermediate species for the formation of hyperbenzene. Figure 3a shows the overview STM image taken after deposition of 0.5 ML DMTP onto the Ag(111) surface at 443 K with a flux of $f$ = 0.5 ML/h. Domains of two different hexagonal species were observed on the surface as labelled by the green and blue frames. The structure of the hexagon with larger cavity was revealed by the zoom-in STM image in Figure 3b. The hexagons contain alternatingly arranged corners and protrusions, which are assigned to MTP units and Ag atoms, respectively, similar to those in the zigzag organometallic chains. In addition, the corner-to-corner distances of the hexagon are 16.0 Å, in agreement with those of the zigzag organometallic chains. Therefore, these hexagons are assigned to organometallic (MTP-Ag)$_6$ macrocycles consisting of six MTP units and six Ag atoms linked by C-Ag-C bonds, as illustrated by the molecular model overlaid in Figure 3b.

Besides the organometallic (MTP-Ag)$_6$ species, hexagons with a lower number of Ag atoms, (MTP)$_6$Ag$_x$ (x<6), were also observed on the sample in Figure 3a. Figure 3c shows an area with a mixture of hyperbenzenes (MTP)$_6$ and two-fold metalated hexagons (MTP)$_6$Ag$_2$ (molecular model overlaid in Figure 3c). The hyperbenzene is marked with a white dotted hexagon. Its identity is verified by the uniform lengths of the six sides of the hexagon motifs, in line with that in Figure 1e. The (MTP)$_6$Ag$_2$, labelled with distorted yellow dotted hexagons in Figure 3c, appears as an elongated hexagon with four short and two long sides. Its structural assignment has been confirmed by the following considerations. First, protrusions were observed in the center of the two long sides. These protrusions are attributed to Ag atoms. In addition, the corner-to-corner distances (15.9 Å) of the longer sides indicate that the involved two MTP units are linked with C-Ag-C bonds, in line with the distances (16.0 Å) in the organometallic chains. Second, the four short sides with corner-to-corner distances of 13.1 Å suggest a direct covalent C-C linkage between the two MTP units.

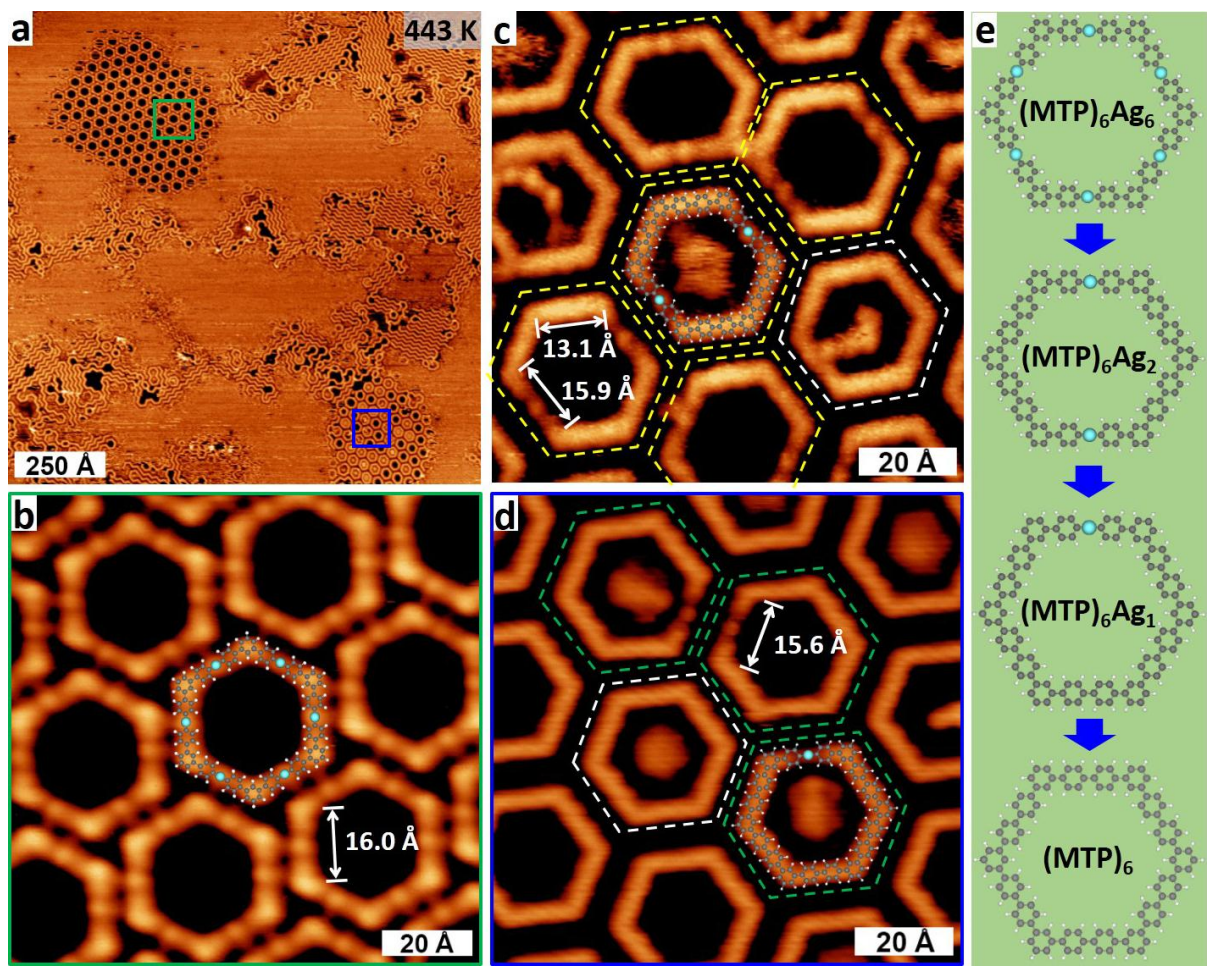

Figure 3. (a) STM image taken after deposition ($f$ = 0.5 ML/h) of 0.5 ML DMTP onto Ag(111) held at 443 K. (b),(d) Zoom-in of the green and blue framed region in panel (a). (c) Section of partially metalated hyperbenzene molecules observed on the sample in panel (a). Tunneling parameters: (a), (b) $U$ = 1.4 V, $I$ = 0.15 nA; (c) $U$ = 0.44 V, $I$ = 0.19 nA; (d) $U$ = 0.88 V, $I$ = 0.17 nA. (e) Scheme for the evolution of cyclic (MTP-Ag)$_6$ to hyperbenzene (MTP)$_6$. Grey spheres represent carbon atoms; white, hydrogen; cyan, silver.

Furthermore, singly metalated hexagons (MTP)$_6$Ag with only one Ag atom were observed, as can be seen in the blue-framed domain (bottom right) in Figure 3a. The magnified STM image (Figure 3d) shows, apart from regular hyperbenzene (example marked with white dotted hexagon), also irregular hexagons (marked by green dotted hexagons) with one long and five short sides. The single long side of the irregular hexagon has a corner-to-corner distance of 15.6 Å, similar to that of two long sides in cyclic (MTP)$_6$Ag$_2$. This strongly implies that the irregular hexagons are (MTP)$_6$Ag, which is proposed to be the final intermediate that leads to the formation of hyperbenzene by elimination of the last remaining Ag atom. Therefore, we propose that the macrocycle formation occurs at the stage of organometallic phase under pseudo-high dilution condition. This reaction results in the initial formation of the organometallic (MTP-Ag)$_6$ macrocycle, which then evolves into the final hyperbenzene by stepwise elimination of Ag atoms as illustrated by the reaction scheme in Figure 3e.

Cyclization reactions in solution can often be treated with the conventional Jacobson-Stockmayer (J-S) model,[34, 35] which describes the yield of cyclic products as a function of the reactant concentrations in high-dilution systems. According to the J-S model, the mechanism for the formation of (MTP-Ag)$_6$ macrocycles can be derived as shown in Scheme 1. Six DMTP monomers initially react with six Ag atoms forming a six-membered (MTP-Ag)$_6$ organometallic chain with five C-Ag-C bridges. This chain then undergoes cyclisation forming the (MTP-Ag)$_6$ macrocycle or grows further into a seven-membered (MTP-Ag)$_7$ chain by addition of another MTP-Ag unit. The competition between the cyclisation and chain growth finally determines the yield ratio of (MTP-Ag)$_6$ macrocycles and (MTP-Ag)$_7$ chains. Since this competition depends strongly on the precursor concentration according to the high-dilution principle, the yield of (MTP-Ag)$_6$ macrocycles can be derived as a function of the precursor deposition rate:

$$\% \,(MTP\text{-}Ag)_6 \text{ cycles} = \frac{1}{1+\dfrac{K_L}{6K_C}f\tau} \times 100 \qquad (1)$$

where $K_C$ and $K_L$ (Scheme 1) denote the rate coefficients for cyclisation and growth of the six-membered $(MTP\text{-}Ag)_6$ chains, respectively (see the part (e) in the SI for a detailed derivation).

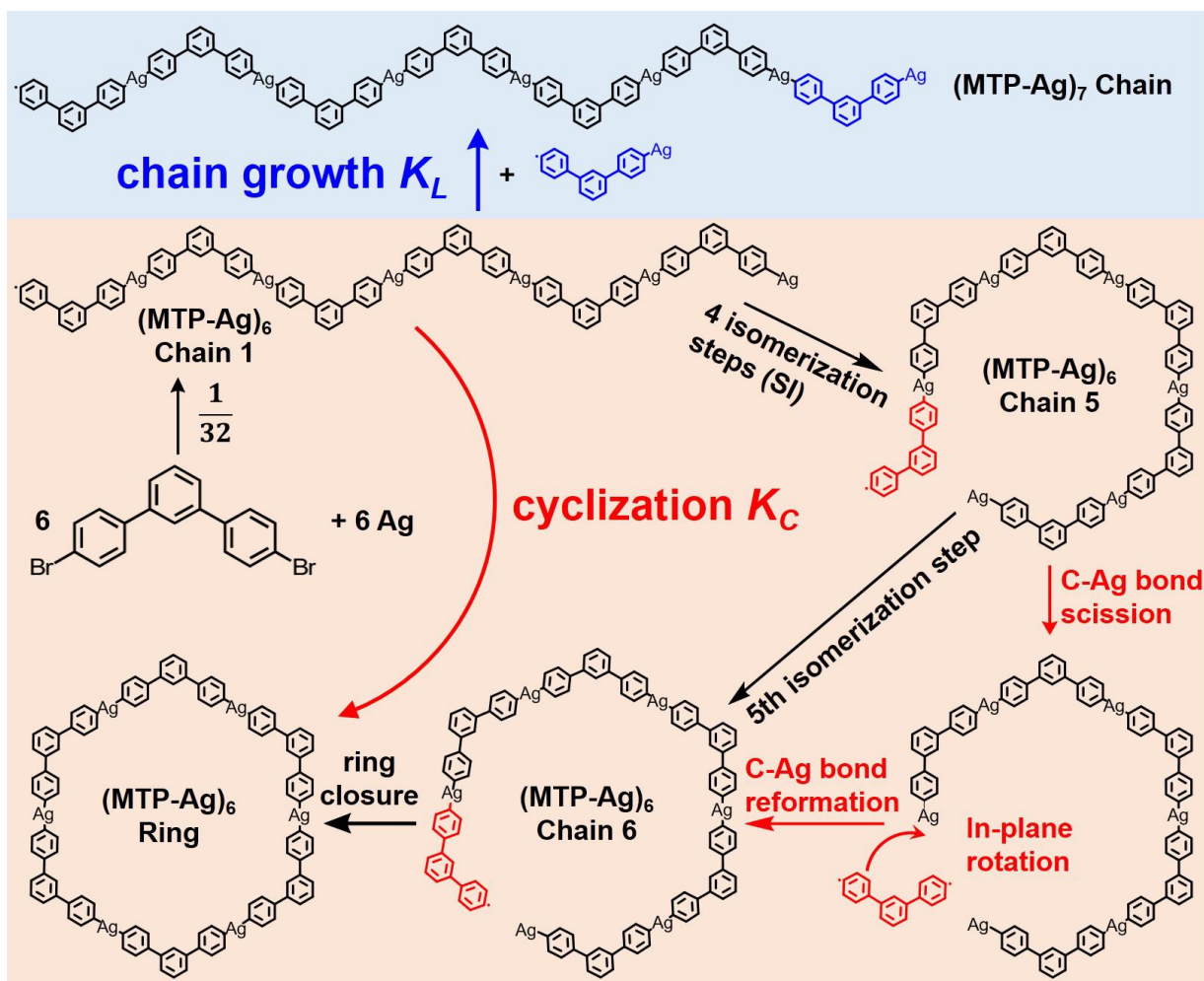

**Scheme 1.** Growth (light blue shaded) and cyclisation (light brown shaded) of all-*trans* configured organometallic $(MTP\text{-}Ag)_6$ Chain 1. The isomerization of organometallic $(MTP\text{-}Ag)_6$ Chain 1 into Chain 6 is achieved by three steps: detachment, in-plane rotation, reattachment of MTP units. $K_C$ and $K_L$ denote the rate coefficients for cyclization and chain growth process, respectively. The details of the 4 isomerization steps from $(MTP\text{-}Ag)_6$ Chain 1 to 5 are shown in Scheme S1 in the supporting information.

For the formation of the (MTP-Ag)$_6$ macrocycle, $K_C$ and $K_L$ are constants depending solely on the mean square end-to-end distance of the six-membered (MTP-Ag)$_6$ chain.[36] $\tau$ is the reaction time for C-Ag-C bond formation, which should be constant at a particular reaction temperature. Considering the observation that hyperbenzene forms from the (MTP-Ag)$_6$ macrocycle, the yield of hyperbenzene should be proportional to the yield of (MTP-Ag)$_6$ macrocycles *i.e.*, it should depend reciprocally on *f*, according to the J-S model and Equation 1. In reality, however, the reciprocal function does not fit to the obtained experimental results well (see the blue curve in Figure 2e). Especially at high deposition rates, the yield of hyperbenzene is higher than that expected by the J-S model.

We attribute the observed deviation from the J-S model to the following differences between on-surface cyclization and cyclization in solution: First, in the J-S model, chains can only grow or remain unchanged, but they cannot become shorter. This is a realistic scenario for solution synthesis with typically direct covalent coupling between the precursor units. On the surface, however, cyclization occurs on the stage of the organometallic species. Since the C-Ag-C bond formation is reversible, chains that have already grown too long for hyperbenzene formation (*i.e.*, (MTP-Ag)$_x$ with x > 6) have a certain probability of dissociation into shorter chains (x ≤ 6) that can still contribute to the formation of hyperbenzene. This is one of the reasons why the yield of hyperbenzene is higher than predicted by the J-S model, especially in the range of higher fluxes, where the chains growth dominates. Second, the (MTP-Ag)$_6$ open chains have with 20 possible configurations due to the random arrangement of the *cis* or *trans* connections between the MTP units (see Scheme S1 in the SI for selected examples). One of these configurations, the all-*cis* Chain 6 (Scheme 1), is the direct conformational precursor for the cyclisation into the (MTP-Ag)$_6$ macrocycle. If the Chain 6 configuration is formed directly from MTP units and Ag, rapid formation of the (MTP-Ag)$_6$ macrocycle is possible without prior isomerization. Since the isomerization requires high-dilution conditions (because otherwise further chain growths is favored), this direct cyclization of Chain 6 increases the macrocycle formation rate compared to the J-S model for high-flux conditions. Note that this effect is caused by the 2D confinement of the chains, *i.e.*, it results from the templating influence of the surface.

For the (MTP-Ag)$_6$ chains with other configurations, *e.g.*, the all-*trans* configured (MTP-Ag)$_6$ Chain 1 (Scheme 1), their cyclisation demands the preceding isomerization into (MTP-Ag)$_6$ Chain 6. This is only realistically possible by the detachment, in-plane rotation, and re-attachment of the MTP units (*e.g.*, from (MTP-Ag)$_6$ Chain 5 to 6, Scheme 1) under the prerequisite of pseudo-high dilution condition. (The alternative flipping of chain segments through 3D space is highly unlikely, because it would require the intermediate desorption of units larger than (MTP)$_2$Ag$_2$, see also the Scheme S2 in the SI for a related all-covalent chain.) Without the detachment/re-attachment mechanism, the chain growth would dominate due to the high probability of attaching an additional MTP unit (light blue shaded region in Scheme 1). Scheme S1 shows the possible six steps for the isomerization of the all-*trans* configured (MTP-Ag)$_6$ chain 1 into the all-*cis* configured Chain 6. In each step, the cleavage of a single C-Ag bond enables the detachment of the MTP units from the chain. Subsequently, the in-plane rotation and recombination of the MTP units eventually result in the isomerization of the (MTP-Ag)$_6$ chains. Noteworthy, once the (MTP-Ag)$_6$ Chain 6 cyclize into the (MTP-Ag)$_6$ macrocycle, the detachment of MTP units is energetically unfavorable compared to that from a chain, because the simultaneous scission of two C-Ag-C bridges is required. This selection process finally leads to the building up of the yield of (MTP-Ag)$_6$ macrocycles.

An additional significant aspect is the template effect of the surface beyond the simple 2D confinement mentioned above. The surface lattice favors the formation of products with matching symmetry. Such template effects have previously been reported for related systems, in particular in comparing the topology of DMTP-derived species on six-fold symmetric Cu(111)[18] and two-fold symmetric Cu(110).[37] Therefore, the formation of six-fold symmetric (MTP-Ag)$_6$ cycle is favored in the initial reaction of DMTP monomers with Ag atoms. This also contributes to the higher than expected hyperbenzene yield at high deposition rates $f$.

**Cyclisation on the stage of covalent species.** Figure 4a, 4b and Figure S4a show STM images taken after deposition of 0.5 ML DMTP ($f$ = 0.5 ML/h) onto the Ag(111) surface held at elevated temperatures of 483, 503, and 523 K. It can be seen that, with increasing substrate temperatures, the yield of hyperbenzene (green region) decreases along with an increase of both yield and lengths of oligophenylene chains (grey regions). Figure 4e shows the corresponding length distributions of the oligophenylene chains formed on Ag(111) held at 463, 483, 503, and 523 K. All the fitted curves shows a Poisson type length distribution, indicating a chain-growth polymerization process of DMTP.[38] In addition, the most probable length of the oligophenylene chains rises with increasing substrate temperature. These phenomena can be interpreted as follows.

The temperature of 463 K represents the threshold for C-C coupling of DMTP on Ag(111). If the reaction is performed at this temperature, there is sufficient time for the cyclisation of the organometallic (MTP-Ag)$_6$ chains, because demetalation of cyclic (MTP-Ag)$_6$ to hyperbenzene is slow. With increasing substrate temperatures, the life-times of the organometallic species become shorter, which undermines the cyclisation process on the stage of organometallic phase. Then, the formation of hyperbenzene proceeds more and more likely through the cyclisation of preformed (MTP)$_6$ oligophenylene chains (Scheme 2) rather than stepwise desilverization of preformed (MTP-Ag)$_6$ macrocycle.

However, due to the irreversibility of the C-C bond formation, the isomerization of oligophenylene chains necessary for cyclisation must follow a different mechanism with higher energy barriers than in the case of the organometallic chains. This leads to a considerably lower probability of cyclisation on the stage of covalent species and hence the lower hyperbenzene yield, as explained in the following: For the organometallic (MTP-Ag)$_6$ chains, their isomerization into an all-*cis* configured chain is achieved by C-Ag bond scission and reformation. This process allows the detached MTP units undergo in-plane rotation (Scheme 1), *i.e.*, the reaction can proceed in 2D confinement. For the (MTP)$_6$ oligophenylene chains, the isomerization occurs exclusively by the flipping of chain moiety (red colored) through 3D space, as illustrated in Scheme 2. This means parts of the (MTP)$_6$ oligophenylene chain desorbs from the surface. Because of this intermediate desorption, hydrocarbons with large molecular weight have the flipping diffusion barriers that are typically higher than the in-plane sliding diffusion barriers.[39] This can easily be understood considering that the flipping and sliding diffusion barriers on a surface relate to the total depth (adsorption energy) and the corrugation of the surface potential, respectively. Estimates of the flipping barriers can be derived from temperature programmed desorption data. For example, the desorption temperature for a submonolayer of *para*-quaterphenyl on Au(111) is around 550 K.[40] Because of the typically larger adsorption energy of phenyl units on Ag(111) than Au(111),[39, 41] the desorption temperature of *para*-quaterphenyl on Ag(111) should be even higher. This means it is highly unlikely that the pentaphenyl moiety (red colored) in the (MTP)$_6$ Chain 4 (Scheme 2) can flip over to form chain 5 on Ag(111) at reaction temperatures below 550 K. Thus, except for the (MTP)$_6$ Chains 5 and 6, for which only small units need to flip, other (MTP)$_6$ chains are completely blocked for cyclisation under the employed reaction temperatures (483 K, 503 K, and 523 K) due to the high barrier for flipping over of moieties bigger than *para*-quaterphenyl. However, the probability of the formation of (MTP)$_6$ Chain 5 and 6 (Scheme 2) by C-C coupling of six MTP units is only $\frac{3}{32}$.

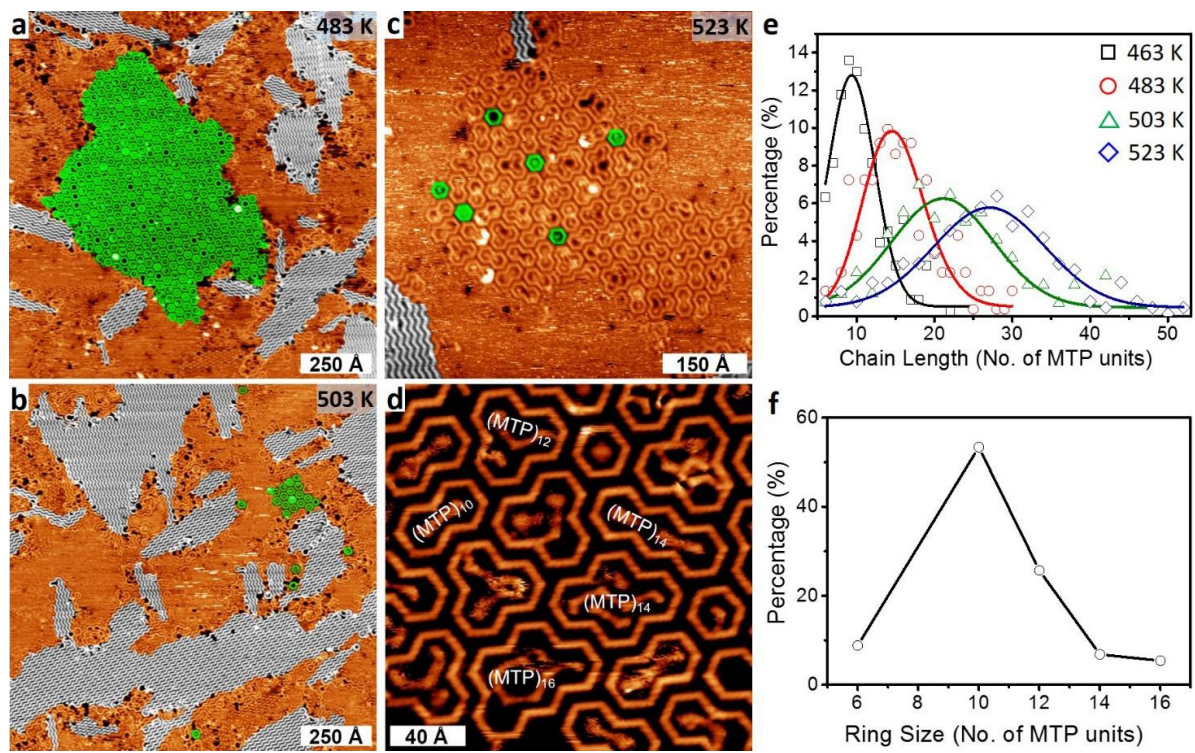

Figure 4. STM images taken at 300 K after deposition ($f$ = 0.5 ML/h) of 0.5 ML DMTP onto Ag(111) held at (a) 483 K, (b) 503 K, and (c) 523 K. The islands of hyperbenzene and zigzag oligophenylene chains were marked out by green and grey shading. (d) Magnified view of a section of the macrocycle species formed on the sample in panel (c). Tunneling parameters: (a) $U$ = 1.6 V, $I$ = 0.12 nA; (b) $U$ = 1.2 V, $I$ = 0.13 nA; (c) $U$ = 0.97 V, $I$ = 0.24 nA; (d) $U$ = 0.76 V, $I$ = 0.12 nA. (e) Length distributions of the zigzag oligophenylene chains on the samples prepared by deposition of 0.5 ML DMTP onto Ag(111) held at 463 (black), 483 (red), 503 (green), and 523 (blue) with Poisson fits. (f) Ring size distributions of the formed macrocycles on the sample in panel (c).

Therefore, even under the pseudo-high dilution condition, the probability of cyclisation on the stage of covalent species is low. This finally leads to the reduced yield of hyperbenzene at high reaction temperatures.

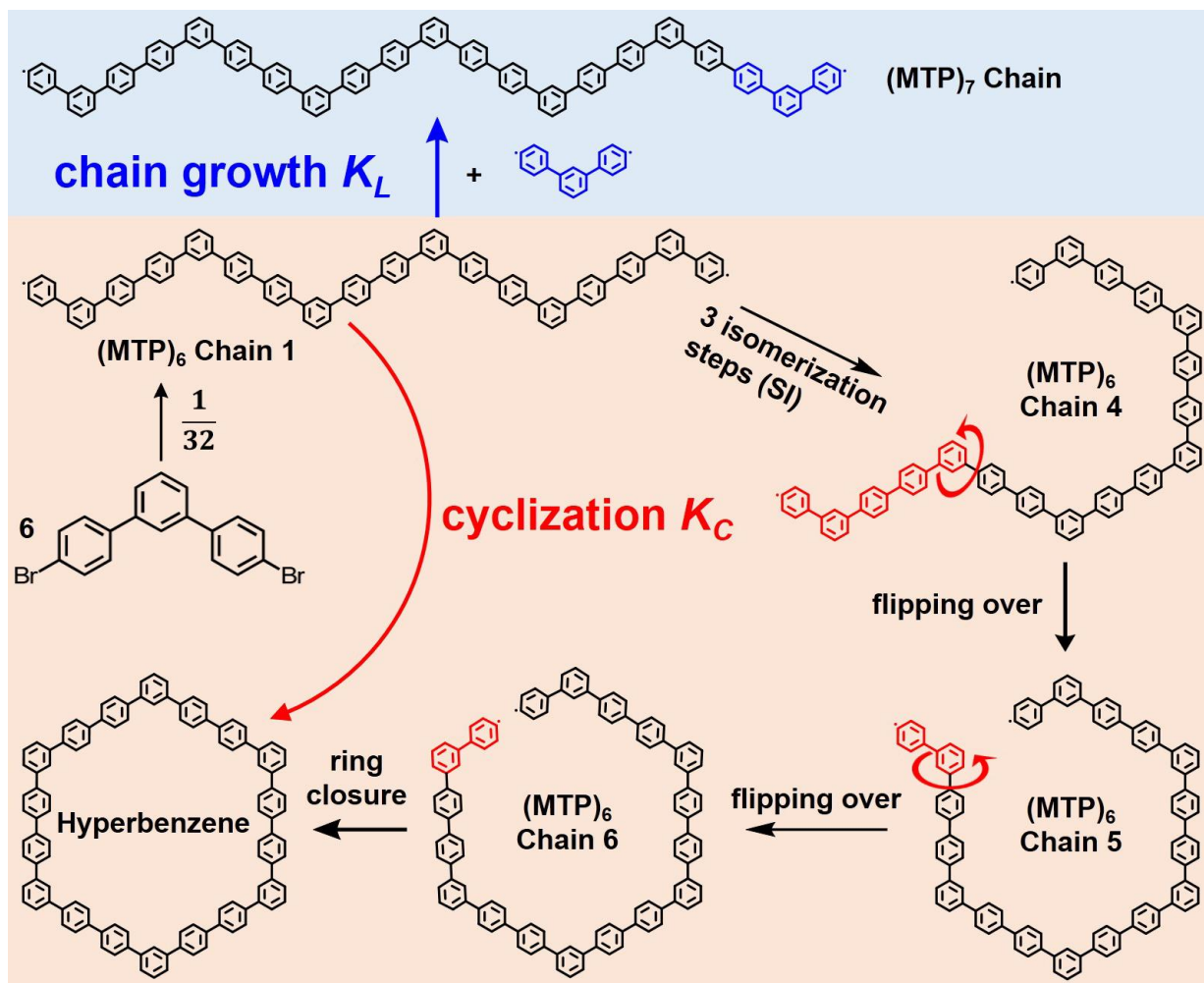

**Scheme 2.** Growth (light blue shaded) and cyclisation (light brown shaded) of all-*trans* configured (MTP)$_6$ oligophenylene chain 1. The isomerization process of (MTP)$_6$ chains necessary for cyclisation is achieved by flipping over of the chain moieties (red colored). $K_C$ and $K_L$ denote the rate coefficients for cyclization and chain growth process, respectively. The details of the 3 isomerization steps from (MTP)$_6$ Chain 1 to 4 are shown in Scheme S2 in the supporting information.

Besides, with the increasing of the substrate temperature, higher sliding diffusion rates of the oligophenylene chains were achieved. This enhanced the probability for the meeting of the ends of two different chains, resulting in the formation of longer chains. Moreover, the formed longer chains cannot dissociate into shorter chains because of the irreversibility of C-C bond formation. As one of the consequences, oligophenylene macrocycles with different sizes are formed at a reaction temperature of 523 K, (Figure 4c), while hyperbenzene (marked green) only appears as a minority species. According to the statistics acquired from the large-scale STM image in Figure S4a, the total yield of macrocycles is around 5%, to which hyperbenzene contributes only 0.5%. The magnified STM image in Figure 4d reveals the detailed structures of the products. Apart from the hexagonal hyperbenzene, the larger macrocycles $(MTP)_{10}$, $(MTP)_{12}$, $(MTP)_{14}$, and $(MTP)_{16}$ were formed, as indicated by the white labels. The size distribution of the macrocycles is given in Figure 4f. As can be seen, the cyclic decamer $(MTP)_{10}$ reaches the highest relative yield. The formation of larger macrocycles can again be attributed to the fact that the lengths of oligophenylene chains ready for cyclization has been enlarged due to the dominance of an irreversible chain growth process at high temperatures, as was discussed above. Noteworthy, no $(MTP)_x$ cycles with x = 1-5, 7-9, 13, and 15 are formed on the surface, probably because these $(MTP)_x$ macrocycles would have ring strain. The major source of overall strain for these macrocycles would be Baeyer strain[42] arising from the deformation of C-C σ bonds between the phenyl units. $(MTP)_x$ cycles with x = 6, 10, 12, 14, and 16 adopt a conformation with less Baeyer strain (no C-C σ bonds deformation) and thus are energetically more favorable to form.

Our observation that cyclisation on the stage of covalent species undermines the yield of hyperbenzene was further supported by otherwise identical experiments on the Au(111) surface as discussed in the following: Figure 5a shows an STM image taken after low-flux deposition ($f$ = 0.5 ML/h) of 0.5 ML DMTP onto Au(111) held at 463 K. The yield of hyperbenzene (green region) is only 9%, which is a surprisingly low value compared to the high yield of 72.5% achieved on Ag(111) at 463 K. Note that the 463 K is again the threshold temperature of C-C coupling on Au(111). This is evidenced by the coexistence of oligophenylene chains and intact DMTP on Au(111) at 443 K (see Figure S5c). The most probable factor accounting for this large yield difference under otherwise identical conditions on Au(111) and Ag(111) is the intrinsically shorter life-time of the organometallic species with C-Au-C bonds comparing to those with C-Ag-C bonds. The short life-time of the C-Au bond is supported by the fact that no stable organometallic species formed from DMTP on the Au(111) surface as confirmed by Figure S5 (and is in agreement with the rarely reported gold-organic intermediates).[43, 44] With the short life-time of C-Au-C bond, cyclisation is more likely to occur on the stage of covalent species, which results in the low cyclisation probability and hence the low yield of hyperbenzene.

Similar to the situation on the Ag(111) surface, increasing the temperature of Au(111) during deposition ($f$ = 0.5 ML/h) of submonolayer of DMTP leads to even more reduced yields of hyperbenzene due to the further decrease of the life-time of the organometallic species. This is manifested as reduced number of hyperbenzene molecules (marked in green) in the STM image series: Figure 5a (463 K), 5b (483 K), 5c (503 K), and Figure S6 (523 K). In addition, the chain length distribution in Figure 5e shows an increase of the most probable lengths of the oligophenylene chains with increasing temperatures of the Au(111) substrate. As can be seen in Figure 5f, the most probable chain length correlates with the substrate temperature on both metal surfaces, and longer chains are observed on Au(111) under otherwise identical conditions. This can be attributed to the higher sliding diffusion rates of the oligophenylene chains achieved on Au(111) than Ag(111) under otherwise identical conditions,[39] which favors the chain growth process. Furthermore, the obtained longer chains have certain probabilities of cyclisation into larger macrocycles. This is supported by the observation of larger covalent oligophenylene macrocycles (marked with cyan in Figure 5d) on Au(111) than Ag(111) at 483 K, including $(MTP)_{10}$ and $(MTP)_{14}$, $(MTP)_{18}$, $(MTP)_{19}$, and $(MTP)_{48}$.

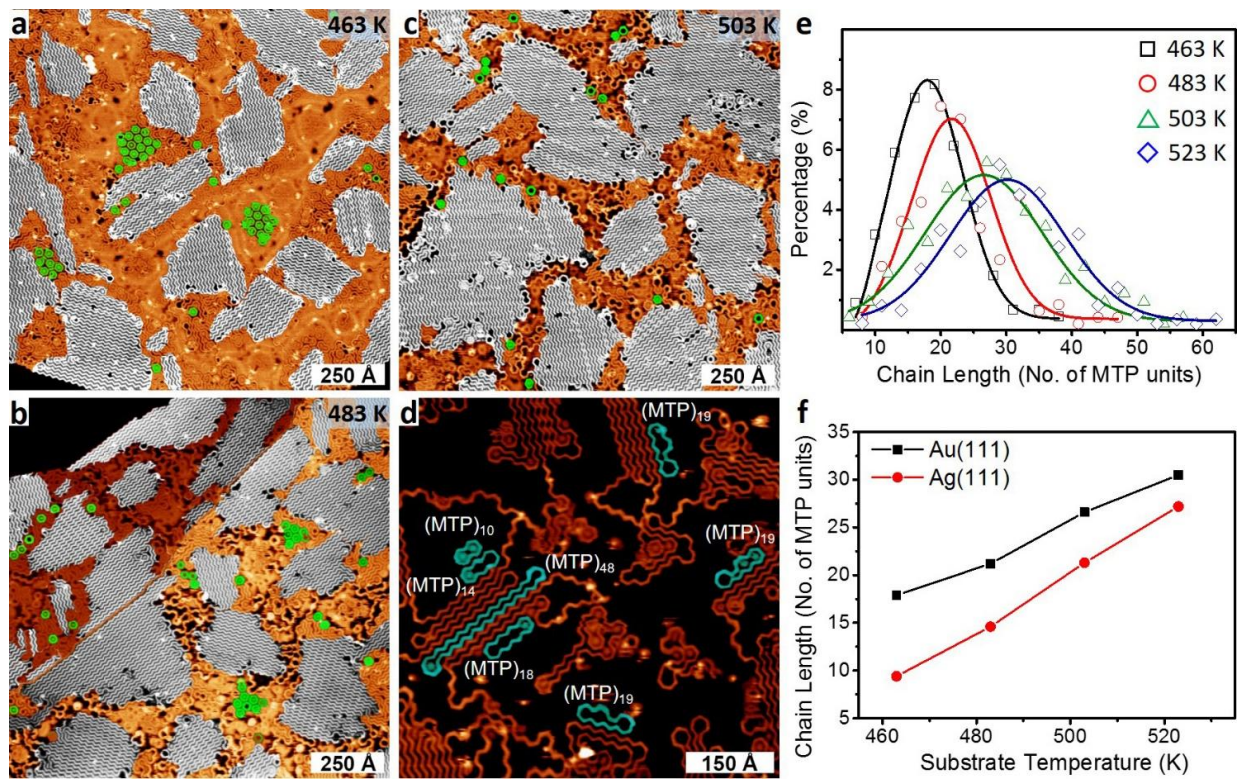

Figure 5. STM images taken after deposition ($f$ = 0.5 ML/h) of 0.5 ML DMTP onto Au(111) held at (a) 463 K, (b) 483 K, and (c) 503 K. The islands of hyperbenzene and zigzag oligophenylene chains were marked out by green and grey shading. (d) A section of the sample in panel (b) with large macrocycles (marked with cyan). Tunneling parameters: (a) $U$ = 1.2 V, $I$ = 0.14 nA; (b), (d) $U$ = 1.8 V, $I$ = 0.10 nA; (c) $U$ = 1.9 V, $I$ = 0.16 nA. (e) Length distributions of zigzag oligophenylene chains on the samples prepared by deposition of submonolayer DMTP onto Au(111) held at 463 K (black), 483 K (red), 503 K (green), and 523 K (blue) with Schulz-Flory and Poisson fits. (f) Dependence curve: most probable lengths of zigzag oligophenylene chains versus substrate temperatures for both Au(111) and Ag(111).

**CONCLUSIONS**

In conclusion, we have experimentally explored the possibilities for (pseudo-)high dilution conditions in macrocycle synthesis on metal surfaces for the first time. By employing an extremely low deposition rate of 0.05 ML/h, we have successfully achieved the high-yield (84%) formation of hyperbenzene ([18]-honeycombene) from DMTP on Ag(111) at 463 K. The mechanism for the formation of hyperbenzene was explored in detail. In the range of the threshold temperature for C-C coupling, e.g. at 463 K, the hyperbenzene is formed majorly by stepwise desilverization of $(MTP-Ag)_6$ macrocycles, which preformed *via* cyclisation of $(MTP-Ag)_6$ chains under pseudo-high dilution conditions. Above the threshold temperature for C-C coupling, the cyclisation process increasingly proceeds by cyclisation of covalent $(MTP)_6$ oligophenylene chains due to the reduced life-time of organometallic species. The cyclisation probability of $(MTP)_6$ oligophenylene chains is considerably lower than that of the $(MTP-Ag)_6$ organometallic chains because of the higher conformational isomerization energy barrier of the oligophenylene precursor chain. Therefore, with the shortening of the life-time of organometallic species, the probability of cyclisation decreases, which finally leads to the reduced yield of hyperbenzene. This observation is further supported by the considerably lower yield of hyperbenzene obtained under otherwise identical conditions on Au(111), in which system the organometallic species with C-Au-C bonds has intrinsically shorter life-time than that with C-Ag-C bonds. Therefore, we conclude that the high yield macrocycle formation on surfaces relies strongly on the efficient cyclisation process on the stage of a reversible (e.g., organometallic) species and a highly diluted precursor. Compared to macrocycle synthesis in solution, the additional reversibility condition as a direct consequence of the 2D confinement makes the detailed understanding of the reaction mechanism and the optimization of the reaction temperature particularly critical.

**EXPERIMENTAL SECTION**

The experiments were performed in two separate UHV systems. The STM measurements were performed in a two-chamber UHV system, which has been described previously,[45] at a background pressure below $10^{-10}$ mbar. The STM probe is a SPECS STM 150 Aarhus with SPECS 260 electronics. All voltages refer to the sample and the images were recorded in constant current mode. Moderate filtering (Gaussian smooth, background subtraction) has been applied for noise reduction. The Ag(111) and Au(111) single crystals with an alignment of better than 0.1° relative to the nominal orientation were purchased from MaTecK, Germany. Preparation of a clean and structurally well controlled Ag(111) and Au(111) surface was achieved by cycles of bombardment with $Ar^+$ ions and annealing at 800 K and 850 K, respectively. As described elsewhere,[18] 4,4''-dibromo-1,1':3',1''-terphenyl (4,4"dibromo-*meta*-terphenyl, DMTP) was made from 4-bromophenylacetylene in a short reaction sequence utilizing a Grubbs-enyne metathesis reaction and a regioselective cobalt-catalyzed Diels-Alder reaction followed by mild oxidation. DMTP was vapor-deposited from a commercial Kentax evaporator with a Ta crucible held at different temperatures for different deposition rates: 329 K (0.05 ML/h), 333 K (0.08 ML/h), 339 K (0.5 ML/h), 367 K (5 ML/h), 389 K (150 ML/h). All STM images were recorded at a sample temperature of 300 K. Coverages were derived from STM images. The DMTP coverage of 1 monolayer (ML) refers to a densely packed layer of zigzag oligophenylene chains and equals 0.052 DMTP molecules per surface Ag atom. The XPS measurements were performed on the Catalysis and Surface Science Endstation located in National Synchrotron Radiation Laboratory (NSRL), Hefei. The detailed description of the endstation can be found elsewhere.[46] The XPS spectra were collected at an emission angle of 60° with respect to the surface normal.

**ASSOCIATED CONTENT**

**Supporting Information**. XP spectra, extra STM images, detailed isomerization schemes, derivation of the equation (1) and stationary concentration of DMTP during its deposition onto 463 K Ag(111) with a flux of $f$ = 0.05 ML/h. This material is available free of charge *via* the Internet at http://pubs.acs.org.

**AUTHOR INFORMATION**

Corresponding Author

* michael.gottfried@chemie.uni-marburg.de

* jfzhu@ustc.edu.cn

Notes

The authors declare no competing financial interest.

**ACKNOWLEDGMENT**

J.F.Z. acknowledges the financial support from the National Basic Research Program of China (2013CB834605), the National Natural Science Foundation of China (21473178 ) and Scientific

Research and Users with Potential Grants of Hefei Science Center of CAS (2015SRG-HSC031, 2015HSC-UP022). J.M.G. thanks the Deutsche Forschungsgemeinschaft for support through SFB 1083 and GO 1812/2-1 as well as the Chinese Academy of Sciences for a Visiting Professorship for Senior International Scientists (Grant No. 2011T2J33). Q.T.F thanks the Alexander von Humboldt-Foundation for Research Fellowship for Postdoctoral Researchers.